\documentclass[manuscript,screen]{acmart}

\AtBeginDocument{%
  \providecommand\BibTeX{{%
    \normalfont B\kern-0.5em{\scshape i\kern-0.25em b}\kern-0.8em\TeX}}}

\DeclareMathOperator{\arccosh}{arccosh}
\newcommand{\iALS}{iALS}
\newcommand{\EASER}{EASE$^R$}
\newcommand{\Cref}[1]{Section~\ref{#1}}
\usepackage{amssymb}
\usepackage{amsthm}
\usepackage{amsfonts}
\usepackage{caption}

\newlength\defaultparindent
\AtBeginDocument{\setlength\defaultparindent{\parindent}}

\usepackage[tight,nooneline]{subfigure}
\copyrightyear{2020} 
\acmYear{2020} 
\setcopyright{acmlicensed}
\acmConference[RecSys '20]{Fourteenth ACM Conference on Recommender Systems}{September 22--26, 2020}{Virtual Event, Brazil}
\acmBooktitle{Fourteenth ACM Conference on Recommender Systems (RecSys '20), September 22--26, 2020, Virtual Event, Brazil}
\acmPrice{15.00}
\acmDOI{10.1145/3383313.3412219}
\acmISBN{978-1-4503-7583-2/20/09}

\begin{document}

\title{Performance of Hyperbolic\,Geometry\,Models on Top-N\,Recommendation\,Tasks}


\author{Leyla Mirvakhabova}
\authornote{These authors contributed equally to the paper.}
\author{Evgeny Frolov}
\authornotemark[1]
\email{evgeny.frolov@skoltech.ru}
\orcid{0000-0003-3679-5311}
\affiliation{%
  \institution{Skolkovo Institute of Science and Technology}
  \city{Moscow}
  \country{Russia}
}

\author{Valentin Khrulkov}
\authornotemark[1]
\additionalaffiliation{%
  \institution{Yandex.Research}
  \city{Moscow}
  \country{Russia}
}
\author{Ivan Oseledets}
\additionalaffiliation{%
  \institution{Institute of Numerical Mathematics, Russian Academy of Sciences}
  \city{Moscow}
  \country{Russia}
}
\affiliation{%
  \institution{Skolkovo Institute of Science and Technology}
  \city{Moscow}
  \country{Russia}
}

\author{Alexander Tuzhilin}
\affiliation{%
  \institution{New York University}
  \city{New York}
  \state{NY}
  \country{USA}
}

\begin{abstract}
We introduce a simple autoencoder based on hyperbolic geometry for solving standard collaborative filtering problem. In contrast to many modern deep learning techniques, we build our solution using only a single hidden layer. Remarkably, even with such a minimalistic approach, we not only outperform the Euclidean counterpart but also achieve a competitive performance with respect to the current state-of-the-art. We additionally explore the effects of space curvature on the quality of hyperbolic models and propose an efficient data-driven method for estimating its optimal value.
\end{abstract}


\maketitle
\renewcommand{\shortauthors}{Mirvakhabova and Frolov, et al.}

\section{Introduction}
Recent reports \cite{basilico2019trends} and meta-studies \cite{dacrema2019worrying,rendle2019difficulty,rendle2020neural} on the performance of collaborative filtering models have revealed that \emph{in the standard scenario} many modern complex algorithms improve upon traditional techniques only marginally, if at all. This observation, in turn, questions the necessity of complex architectures and overall progress of the field.

At the same time, several recently introduced techniques \cite{nikolakopoulos2019recwalk,nikolakopoulos2019personalized} show a much more pronounced improvement over common baselines without involving any complex machinery of deep learning and only utilizing random walks on graphs. One of the key components introduced by the authors is a better prior assumption on the nature of data, which depends on the adequate representation of relations between items. The authors coin the term \emph{item model} for it. As it follows from their experiments, a proper item model may significantly improve the performance of the final solution. Hence, \emph{capturing an underlying nature of data} becomes a \emph{driving factor for improving recommendations quality}.



Inspired by this remarkable result, we aim to explore alternative ways of modeling data. Instead of seeking for a good item model in Euclidean space, we propose to translate an entire solution to space with more suitable geometry, where the structure of a given data will find more natural representation. We rely on the fact that typical user-item interactions data form a bipartite graph, which in turn renders a so-called complex network \cite{guillaume2006bipartite}. Properties of complex networks have been extensively studied before and are known to be tightly connected to a hyperbolic geometry \cite{krioukov2010hyperbolic}. This fundamental result gave rise to a large body of work with practical applications in various fields from natural language and image processing \cite{nickel2017poincare,khrulkov2019hyperbolic} to recommender systems \cite{vinh2018hyperbolic,chamberlain2019scalable}.

In this work, we tackle the problem of hyperbolic recommender systems from a slightly different perspective. We start from minimal possible architecture design and introduce a non-euclidean autoencoder with a single hidden layer. It operates directly on an item space and translates it to a hyperbolic geometry, which enables capturing an underlying structure of data more efficiently. It does not require any specific notion of users other than simply a ``bag of items''. Remarkably, such a minimalistic approach does not limit the predictive ability of our model and allows not only outperforming Euclidean analogs with identical architecture but also achieving a competitive quality in comparison to recent and architecturally more complex solutions. To the best of our knowledge, this is the first application of autoencoders on hyperbolic geometry to the collaborative filtering task. 


We additionally explore the concept of \emph{space curvature} to better model structural properties of data. The choice of curvature plays an important role in achieving better data representation. Previously, it was either simply neglected (by utilizing a default value) or treated as a hyperparameter, which optimal value was estimated using some grid-search pipeline \cite{vinh2018hyperbolic}.
In this work, we eliminate the need for extensive search by providing a practical algorithm for  \emph{estimating the corresponding space curvature factor directly from the geometry of input data}. Our experimental results confirm an improvement over the default option.
We share the entire source code for reproducing our work on Github\footnote{https://github.com/evfro/HyperbolicRecommenders}.





\section{Related work}
Hyperbolic geometry has already been applied to the recommender systems domain in several recent studies. The first paper proposing to utilize hyperbolic geometry \cite{vinh2018hyperbolic} adapted a gyrovector formalism to model the user-item relations by performing metric learning in hyperbolic spaces. Additionally, the authors analyzed the effect of different space curvatures on model performance through grid search. The authors of the paper \cite{chamberlain2019scalable} continued this line of research with an emphasis made on large scale data. One of the major ideas there is representing users not as separate entities but rather as the average of their items representations computed via hyperbolic midpoint. 
The paper \cite{schmeier2019music} is very similar in spirit to the previously discussed papers. The only significant difference, which constitutes the major contribution of the work, is that the authors utilize domain knowledge for calculating Bayesian priors that are used to construct the links between entities (for instance, between global objects such as radio stations and smaller objects such as artists). 

Relevant to our work as well, considerable progress has been made using autoencoder models in application to recommender systems. The paper \cite{sedhain2015autorec} was among the first works incorporating standard autoencoder frameworks to collaborative filtering. More recent architectures such as MultVAE \cite{liang2018variational} and RecVAE \cite{shenbin2020recvae} stepped further by moving towards variational autoencoders and introducing specific loss functions tailored to the task of collaborative filtering. As we show further, the accompanying increase of complexity in models' architecture may not be necessary when more appropriate assumptions about data geometry are made.


\section{Reminder on hyperbolic geometry}
Let us firstly provide a gentle introduction to existing models of hyperbolic geometry, and describe a method that allows estimating a ``degree'' of hyperbolicity in a given dataset.
\subsection{Hyperbolic spaces}
\paragraph*{The Poincar\'e ball model}\label{sec:poincare}
Formally, $n$-dimensional hyperbolic space is a manifold of constant negative curvature. This property makes it analogous to the ordinary Euclidean sphere, which has constant positive curvature; however, the geometrical properties of hyperbolic spaces are very different. In order to work with these spaces, several equivalent \emph{models} exist. We use the Poincar\'e ball model $\mathcal{P}^n$ due to its relative conceptual simplicity.
It is realized by the $n$-dimensional open unit ball $\mathcal{B}^n = \{\mathbf{x} \in \mathbb{R}^n \colon  \|\mathbf{x}\| < 1\}$.
The open unit ball is equipped with the Riemannian metric tensor 
\begin{equation}
  g_{\mathcal{P}}(\mathbf{x}) = \lambda_x^2 g^E,
\end{equation}
where $\lambda_x = \frac{2}{1-\|x\|^2}$ -- conformal factor and $g^E$ denotes the Euclidean metric tensor (the identity matrix). We emphasize on the fact that \emph{a mapping to Poincar\'e model from Euclidean space is conformal} -- i.e., this mapping preserves the angle between vectors.
Here and throughout this paper, by $\|\cdot\|$, we denote the ordinary Euclidean norm. 

For two points $\mathbf{x}, \mathbf{y} \in \mathcal{B}^n$, the distance $d_{\mathcal{P}}(\mathbf{x},\mathbf{y})$ in the Poincar\'e model is defined in the following manner:
\begin{equation}{\label{pdist}}
	d_{\mathcal{P}}(\mathbf{x},\mathbf{y}) = \arccosh \Big(1 + 2 \frac{\|\mathbf{x}-\mathbf{y}\|^2}{(1-\|\mathbf{x}\|^2)(1-\|\mathbf{y}\|^2)} \Big).
\end{equation}  

If we take a closer look at the distance definition \eqref{pdist}, we may notice that for points close in Euclidean distance, the hyperbolic distance in the Poincar\'e model may be exponentially large. This happens, for instance, in the case of neighboring points located near the boundary, and denominator in \eqref{pdist} tends to infinity. As was established in \cite{krioukov2010hyperbolic}, this property makes hyperbolic spaces suitable for embeddings of \emph{complex networks}. Roughly speaking, the exponential increase of volumes in hyperbolic spaces is analogous to defining characteristics of complex networks, such as the power-law node degree distribution. This property readily emerges in recommender systems, where the user--item networks typically exhibit similar properties, e.g., the power-law distribution of user/item popularity \cite{cano2006topology}. 

\paragraph*{The Lorentz model}\label{sec:lorentz}
In the part of our experiments, we switch to the Lorentz model of hyperbolic geometry. This model realizes the hyperbolic space $\mathbb{H}^n$ as a hyperboloid in $\mathbb{R}^{n+1}$. In order to specify it, it is convenient to introduce the following Minkowski inner product
$\langle \mathbf{x}, \mathbf{y} \rangle_{\mathcal{L}} = \sum_{i=1}^{n} x_i y_i - x_{n+1} y_{n+1}.$
Then, the hyperbolic space $\mathbb{H}^n$ is specified by the set $\mathcal{L}^n$, defined as $\mathcal{L}^n = \{ \mathbf{x} \in \mathbb{R}^{n+1} \colon \langle \mathbf{x}, \mathbf{x} \rangle_{\mathcal{L}} = -1, x_{n+1} > 0 \}$ endowed with the Riemannian metric tensor $g_{\mathcal{L}}$ given by
$g_{\mathcal{L}}(\mathbf{x}) = \mathrm{diag} (1, 1, \ldots, 1, -1).$
This metric induces the following \emph{hyperbolic} distance between two points on $\mathcal{L}^n$:
\begin{equation}\label{pdist_hyper}
    d_{\mathcal{L}}(\mathbf{x}, \mathbf{y}) = \arccosh (-\langle \mathbf{x}, \mathbf{y} \rangle_{\mathcal{L}}).
\end{equation}

\subsection{Hyperbolic neural networks}\label{sec:hypnn}
Further in this section, we briefly discuss the hyperbolic functions and layers forming hyperbolic neural networks \cite{ganea2018hyperbolic}.
Since hyperbolic spaces are not vector spaces, one has to decide on how to translate common operations utilized in machine learning models without violating the structure of a new space. In this paper we incorporate a framework of M\"obius \emph{gyrovector spaces} \cite{ungar2008gyrovector}, widely used in practice \cite{vinh2018hyperbolic, ganea2018hyperbolic}. This formalism for hyperbolic geometry conveniently provides an analogy to how vector spaces are used in Euclidean geometry.

We additionally introduce a new hyperparameter $c$, inversely related to the curvature of the hyperbolic space ($K = -\frac{1}{c^2}$). The corresponding Poincar\'e ball is defined as follows: \mbox{$\mathcal{B}_c^n = \{\mathbf{x} \in \mathbb{R}^n: c\|\mathbf{x}\|^2 <1, c \geq 0\}$}. The conformal factor is then modified as $\lambda_{\mathbf{x}}^c = \frac{2}{1 - c\|\mathbf{x}\|^2}$. In practice, the choice of $c$ allows balancing between hyperbolic and Euclidean geometries, which is made precise by the fact that with $c \to 0$, all the formulae take their usual Euclidean form. Most of the previous attempts to adapt hyperbolic geometry for applications in recommender systems have ignored this parameter and used the default value of $c=1$. The authors of \cite{vinh2018hyperbolic}, however, reported a relatively large variation in quality with different curvatures and performed a grid-search for identifying its optimal value. In this work, we explore an alternative way and provide a simple technique to pre-estimate curvature from data (described later in this section).

We also utilized the following operations, firstly introduced to the machine learning community in \cite{ganea2018hyperbolic} and \cite{nickel2017poincare}.
\begin{itemize}
    \item \textbf{Exponential and logarithmic maps}. These operations allow us to transition between a hyperbolic point and its corresponding Euclidean parameterization.
    \item \textbf{Hyperbolic linear layer.} This layer generalizes the standard feedforward linear layer, i.e., matrix multiplication followed by the addition of a bias vector, using so-called M\"obius addition. 
\end{itemize}
The exact formulas used in the aforementioned layers are quite cumbersome, and we omit them for the sake of concise description. For a thorough review, we refer the reader to \cite{ganea2018hyperbolic}.

There are two ways to parametrize bias parameters in hyperbolic linear layers: the first one is to treat bias as a parameter on \emph{hyperbolic space} and optimize it via Riemannian optimization \cite{bonnabel2013stochastic,becigneul2018riemannian}, the second is to keep it as Euclidean parameter and map onto hyperbolic space during calculations. In our experiments, we use both variants; resulting layers are termed \texttt{M\"obiusLinear} and \texttt{HypLinear} respectively. As the optimization algorithm, we used the recently proposed Riemannian Adam \cite{becigneul2018riemannian}.  To ensure numerical stability, we apply clipping by norm for points in the Poincar\'e ball, constraining the norm not to exceed $\frac{1}{\sqrt{c}} (1 - 10^{-3})$.

The curvature of the underlying data manifold is often neglected in the literature on hyperbolic deep learning. However, it is an informative parameter as its value is closely related to the optimal radius of Poincar\'e ball (recall that the curvature of $\mathcal{B}_c^n$ is given by $-\frac{1}{c^2}$). Hence, instead of using a unit Poincar\'e ball, a more practical approach would be to \emph{estimate} a suitable radius value (or, equivalently, the value of $c$) that gives a better data representation. An estimate of this value can be obtained with the method introduced in \cite{khrulkov2019hyperbolic}. The method is based on the concept of Gromov's $\delta$-hyperbolicity \cite{gromov1987hyperbolic}. Informally, space is $\delta$-hyperbolic if metric relations between four points in such a space are the same in a tree (in the sense of graph theory) up to a constant $\delta$. Trees are $0$-hyperbolic spaces and serve as a reference in this definition. $\delta$-Hyperbolicity is defined for an arbitrary metric space, even discrete. Given the value of $\delta$, the value of $c$ can be approximated as $c(X)=\left(\frac{0.144}{\delta}\right)^{2}.$
To estimate the $\delta$ value itself, we use the algorithm proposed in \cite{fournier2015computing}. 

From a practical perspective, we propose to estimate the curvature in the following manner. The first step is to obtain compact representations via truncated singular value decomposition $USV^\top$ of the user-item interaction matrix. Then, one may estimate the $\delta$ values using any of the components $U$ and $V$ as well as $US$ and $VS$. We base our calculations on $VS$. Our preliminary experiments have indicated that it gives the closest match to a grid-search-based estimation.

Due to the computational complexity of the algorithm for $\delta$ approximations, it is common to perform the computations on randomly chosen smaller subsets of data, and average the obtained results across several trials. This characteristic of data manifold is quite stable; thus one does not need to consider large rank values for truncated SVD as well as performing numerous trials; in our experiments, we concluded that it suffices to use the rank equal to $100$ and sample a submatrix of size $1500 \times 1500$ across $10$ trials.


\subsection{Hyperbolic VAE}\label{subsec:hvae}
Our starting experiments with autoencoders yielded promising results, so we decided to consider a more complicated architecture, namely, hyperbolic variational autoencoder -- a generalization of its Euclidean counterpart \cite{nagano2019differentiable, Skopek2020Mixed}. Similarly to previously described hyperbolic models, hyperbolic VAE reduces to Euclidean model when curvature tends to zero. The most intriguing part of this architecture is a generalization of the normal distribution to Riemannian manifolds. Informally, the idea behind sampling in Riemannian manifolds is the following: one can sample from tangent space (the tangent space of hyperbolic space is isometric to Euclidean) and then map the sampled point to the manifold using exponential mapping (see Figure \ref{fig:hyp_normal} for an example). This method allows efficient sampling from distribution on hyperbolic spaces as well as efficient computation of log-probability of the sample; the parametrization of the density function is differentiable and straightforward to compute.\par

\vspace{5.0\lineskip}
\noindent
\begin{minipage}[h]{0.75\textwidth}
\setlength{\parindent}{\defaultparindent}
The structure of the hyperbolic VAE is thus the following: it consists of encoder and decoder with \emph{usual} layers; since the parameters itself are Euclidean, we can use standard gradient-based optimization. The output of the last layer of the encoder is mapped to hyperbolic space via exponential mapping, and the sampled latent vector is mapped back to Euclidean space by logarithmic mapping and plugged into decoder layers. 

The authors of the paper proposing hyperbolic VAEs suggest utilizing the Lorentz (hyperboloid) model of hyperbolic geometry (Section \ref{sec:lorentz}) as variational sampling requires more stable and efficient computations. Since all models of hyperbolic space are \emph{equivalent} it is common to switch from one model to another. In our experiments with VAEs, we have similarly used the Lorentz model of hyperbolic geometry.
\end{minipage}
\hfill%
\begin{minipage}[t]{0.23\textwidth}
    \centering
    \includegraphics[width=0.7\textwidth]{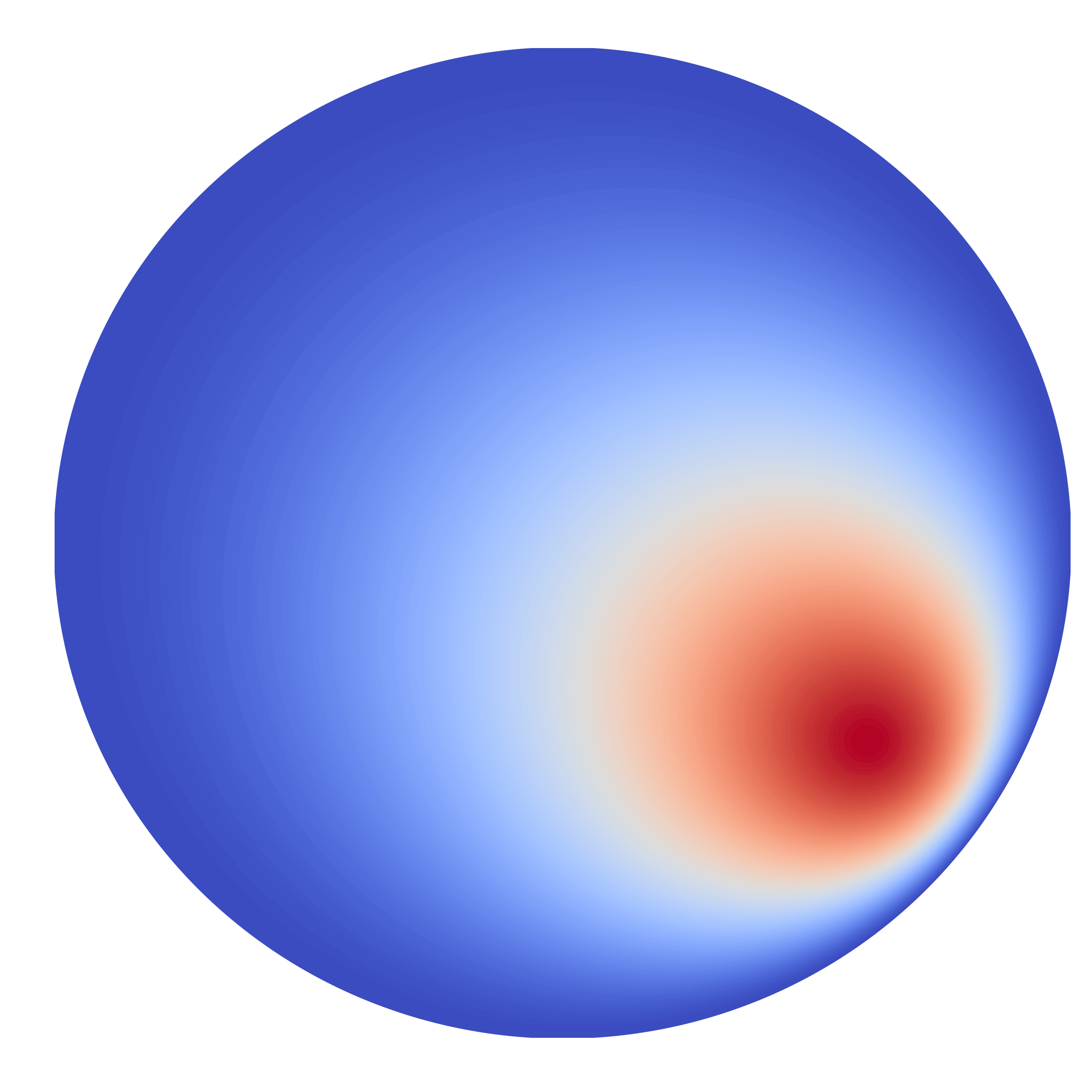}
    \captionof{figure}{An example of a normal distribution on a unit Poincar\'e ball with mean $(0.6, -0.4)$ and standard deviation $2$.}
    \label{fig:hyp_normal}
    \Description[Visual representation of variational sampling on a Poincar\'e ball]{<Figure shows projection onto a disk. We observe concentric structure on the disk with the color gradient from the center outwards, which corresponds to how normal distribution fades towards its tails in 2D.>}
\end{minipage}


\section{Experiments}

We conduct two groups of experiments. In the first group, we perform a simple \emph{ablation study} by comparing a standard autoencoder based on Euclidean geometry with its two hyperbolic counterparts. All three models have an identical architecture with a single hidden layer. The hyperbolic models only differ in the way the hyperbolic linear layer is implemented, as we describe in Section \ref{sec:hypnn}. All models target the same optimization objective based on a binary cross-entropy loss and are tuned using the same grid of hyperparameters.

In the second group of experiments we implement a \emph{variational extension} of our hyperbolic autoencoder as described in Section \ref{subsec:hvae}. The single-layer architecture is preserved. We compare this implementation with various state-of-the-art models, using the results from \cite{dacrema2019troubling} as a reference.
In all experiments, in order to find an optimal configuration, we let models run 40 trials with a 20 epochs budget for each trial on a validation set. We additionally utilize Bayesian search in the hyperparameter space, which is commonly preferred in practice over the random search. 

We have carefully followed methodological principles and practical recommendations outlined in \cite{dacrema2019troubling,dacrema2019worrying} in order to avoid common mistakes and issues.
We used the source code that accompanies \cite{dacrema2019troubling} for generating the data splits for training, validation, and test phases.  We have selected four publicly available datasets that are commonly used for evaluation of recommendation models: \emph{Movielens-1M}, \emph{Pinterest}, \emph{Movielens-20M}, and \emph{Netflix}.

All four datasets are present in \cite{dacrema2019troubling}. The first two are described in the section devoted to the NCF model \cite{he2017neural} and are used to evaluate algorithms in a basic leave-one-out scenario on known users, which only measures weak generalization. On these datasets, we target \text{NDCG}$@10$ for finding optimal configuration.
The other two datasets correspond to the section devoted to MultVAE \cite{liang2018variational} and are designed to measure strong generalization by ensuring that no test users are present in the training data. We target \text{NDCG}$@100$ for finding optimal configuration as in the original MultVAE paper.

\begin{figure*}
    \centering
    \subfigure[][]{%
        \label{fig:urm_comp}
        \includegraphics[width=\textwidth]{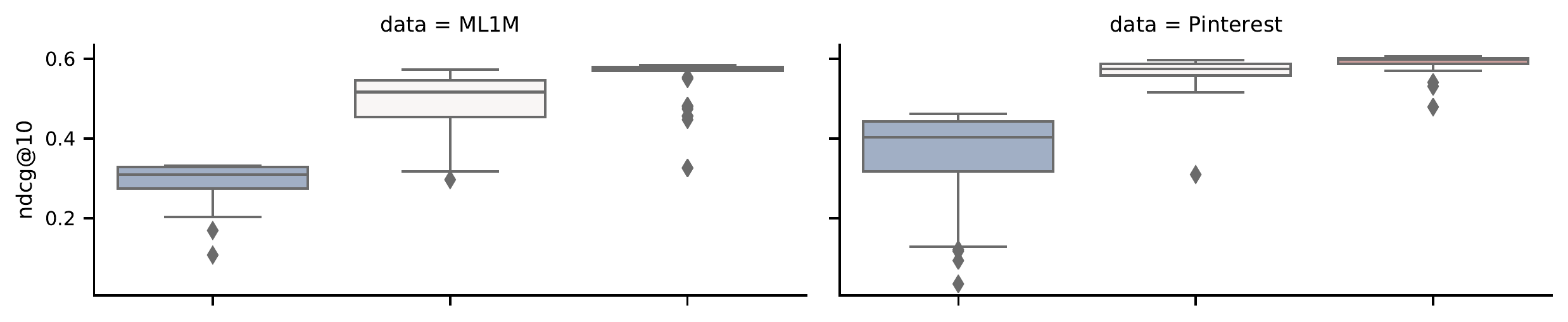}}
    \hfill%
    \subfigure[][]{%
        \label{fig:batch_comp}
        \includegraphics[width=\textwidth]{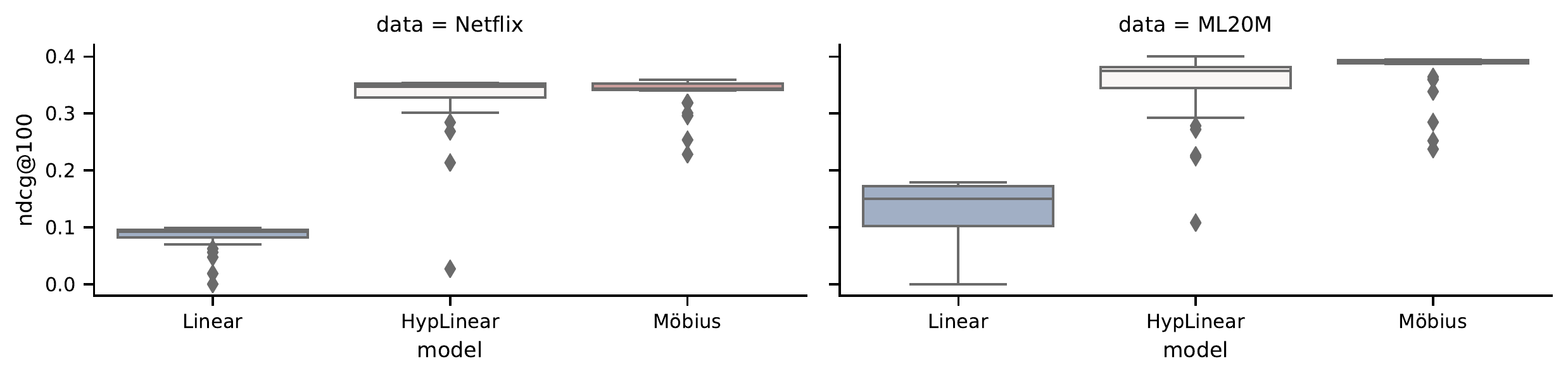}}
    \caption{Comparison of the Euclidean and hyperbolic models based on 40 validation runs for \subref{fig:urm_comp} NCF and \subref{fig:batch_comp} MultVAE experiments.}
    \label{fig:hyp_comp}
    \Description[Comparing Hyperbolic autoencoders against Euclidean.]{<In all experiments we observe a significant boost in quality when transitioning from standard Euclidean autoencoder to the one based on hyperbolic geometry. Between the two hyperbolic models, the M\"obius model is always slightly better.>}
\end{figure*}

\begin{figure*}[t]
    \centering
    \includegraphics[width=\textwidth]{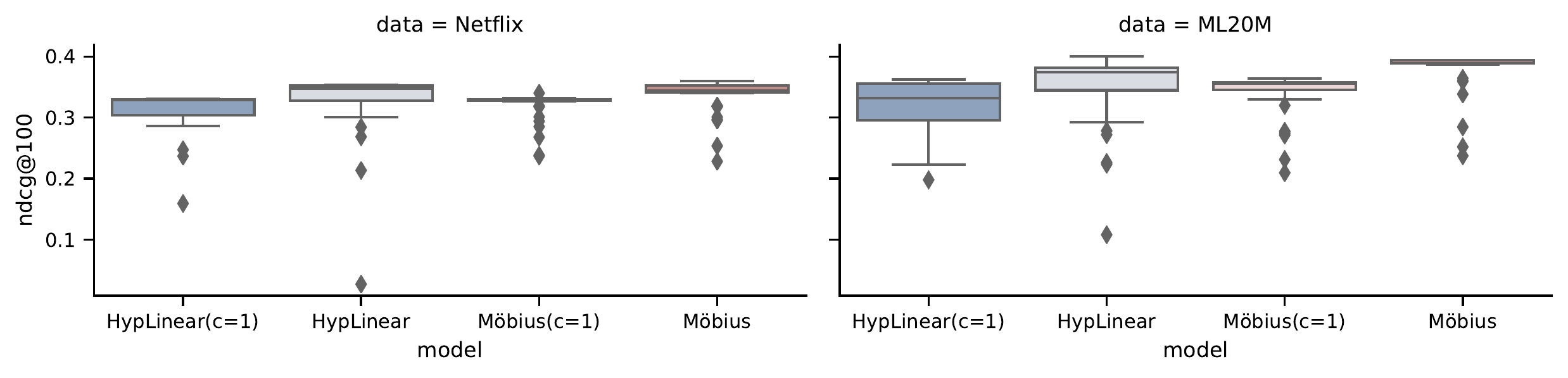}
    \caption{The effect of pre-estimating the value of $c$ on the learning abiltity of hyperbolic autoencoders on the MultVAE data. The results on other datasets are omitted for brevity.}
    \label{fig:batch}
    \Description[Space curvature effect on quality.]{<The proposed technique of pre-estimating values of parameter $c$ allows to improve the quality of hyperbolic models without the need to perform grid search. We observe this effect across all datasets and for both hyperbolic models.>}
\end{figure*}

\begin{table}[ht]
    \caption{Results for the Netflix Prize (left) and Movielens-20M (right) datasets corresponding to the MultVAE experiments. AE denotes the standard Euclidean autoencoder, HAE(H, $c_0$) and HAE(M, $c_0$) denote the hyperbolic autoencoder with \texttt{HypLinear} and \texttt{M\"obiusLinear} layers respectively, trained with $c=c_0$. H-VAE($c_0$) denotes the hyperbolic variational autoencoder trained with $c=c_0$. The best result is marked in bold font, and the second-best result is underlined.}
    \label{tab:Mult-VAE-results}
    \centering
    \footnotesize
    \hspace*{1cm}\begin{tabular}{lcccc}
        \toprule
        & \multicolumn{2}{c}{@50} & \multicolumn{2}{c}{@100} \\
    	& REC 	& NDCG 	& REC 	& NDCG 	\\
        \midrule
        \EASER 		&0.3801	&0.2978	&0.5072	&0.3510	\\
        SLIM 		&\underline{0.4002}	&\textbf{0.3203}	&0.5299	& \underline{0.3752}	\\
        PureSVD		&0.3593	&0.2840	&0.4784	&0.3342	\\
        \iALS		&0.3138	&0.2410	&0.4216	&0.2862	\\
        MultVAE		&\textbf{0.4127} & \underline{0.3167} & \underline{0.5456} &0.3730	\\
        \midrule
        AE & 0.1994 & 0.1333 & 0.3176 & 0.1744\\
        HAE(H, $1$) & 0.3197&	0.2623&	0.4695&	0.3282\\
        HAE(H, $0.004$)&	0.3498&	0.2801&	0.5149&	0.3516\\
        HAE(M, $1$) &	0.3278&	0.2694&	0.4823&	0.3375\\
        HAE(M, $0.004)$ &	0.3565&	0.2872&	0.5211&	0.3586 \\
        \midrule
        \textbf{H-VAE}($0.004$) &0.3793&	0.3122&	\textbf{0.5526}&	\textbf{0.3891} \\
        \bottomrule
    \end{tabular}\hfill%
    \begin{tabular}{lcccc}
        \toprule
        &  \multicolumn{2}{c}{@50} & \multicolumn{2}{c}{@100} \\
        & REC 	& NDCG 	& REC 	& NDCG 	\\
        \midrule
        \EASER 	&0.4608	&0.3267	&0.5860	&0.3711	\\
        SLIM 	& \underline{0.4893}	& \underline{0.3576}	&0.6110	&0.4017	\\
        PureSVD	&0.4371	&0.3117	&0.5544	&0.3538	\\
        \iALS	&0.4406	&0.3090	&0.5631	&0.3521	\\   
        MultVAE &\textbf{0.5222}	&\textbf{0.3690}	& \underline{0.6517}	& \underline{0.4158}	\\
        \midrule
        AE & 0.1994 & 0.1333 & 0.3176 & 0.1744\\
        HAE(H, $1$) & 	0.3725	&0.2780	&0.5477	&0.3521\\
        HAE(H, $0.005$) & 0.4186	&0.3082	&0.6138	&0.3904\\
        HAE(M, $1$) &0.3776	&0.2776	&0.5609	&0.3575\\
        HAE(M, $0.005)$ &0.4148&	0.3062&	0.6055&	0.3874 \\
        \midrule
        \textbf{H-VAE}($0.005$) &0.4500	&0.3407	&\textbf{0.6534}	&\textbf{0.4279} \\
        \bottomrule
    \end{tabular}\hspace*{1cm}
\end{table}
\begingroup
\setlength{\tabcolsep}{3.5pt}

\begin{table}[ht]
    \caption{Results for the Pinterest (left) and Movielens-1M (right) datasets corresponding to the NCF experiments. AE denotes the standard Euclidean autoencoder, HAE(H, $c_0$) and HAE(M, $c_0$) denote the hyperbolic autoencoder with \texttt{HypLinear} and \texttt{M\"obiusLinear} layers respectively, trained with $c=c_0$. The best result is marked in bold font, and the second-best result is underlined.}   
    \label{tab:NCF-results}
    \centering
    \footnotesize
        \begin{tabular}{lcccccc}
        \toprule
        &  \multicolumn{2}{c}{@1} & \multicolumn{2}{c}{@5} & \multicolumn{2}{c}{@10} \\
    	&  HR 	& NDCG 	& HR 	& NDCG 	& HR 	& NDCG 	\\
    	\midrule
    	\EASER 	 & 0.2889 & 0.2889 &0.7053	&0.5057	&0.8682	&0.5589	\\
        SLIM     & 0.2913 & 0.2913 &0.7059	&\underline{0.5072}	&0.8679	&\underline{0.5601}	\\
        PureSVD	 & 0.2630 & 0.2630 &0.6628	&0.4706	&0.8268	&0.5241	\\
        \iALS	& 0.2811 & 0.2811 & \textbf{0.7144}	&0.5061	&\underline{0.8761}	&0.5590	\\   
        NMF     & 0.2307 & 0.2307 & 0.6445 & 0.4434 & 0.8343 & 0.5052\\
        NeuMF  & 0.2801 & 0.2801 & 0.7101 & 0.5029 & \textbf{0.8777} & 0.5576\\
    	\midrule 
    	AE & 0.2789 & 0.1825 & 0.5414 & 0.3659 & 0.7441 & 0.4317\\
        HAE(H, $1$) & 0.2778 & 0.2778 & 0.6847 & 0.4902 & 0.8337 & 0.5390\\
        HAE(M, $1$) & 0.2789 &0.2789 & 0.6884 & 0.4928 & 0.8315 & 0.5397\\
        HAE(H, $0.04$) &\underline{0.2946} & \underline{0.2946} & 0.6806 & 0.4977 & 0.8038 & 0.5381 \\
        HAE(M, $0.04$) & 0.2857 & 0.2857 & 0.6988 & 0.5015 & 0.8466 & 0.5500 \\
        \midrule
        \textbf{H-VAE}($0.04$) & \textbf{0.2957} & \textbf{0.2957} & \underline{0.7126} & \textbf{0.5123} & 0.8760 & \textbf{0.5665}\\
        \bottomrule
    \end{tabular}\hfill%
    \begin{tabular}{lcccccc}
        \toprule
        &  \multicolumn{2}{c}{@1} & \multicolumn{2}{c}{@5} & \multicolumn{2}{c}{@10} \\
    	&  HR 	& NDCG 	& HR 	& NDCG 	& HR 	& NDCG 	\\
    	\midrule
    	\EASER 	&0.2119 & 0.2119	&0.5502	&0.3857 &0.7098 & 0.4374 \\
        SLIM 	&\textbf{0.2207}	&\textbf{0.2207}	&\textbf{0.5576}	& \textbf{0.3953}	& \textbf{0.7162} & \textbf{0.4468} \\
        PureSVD	&0.2132	&0.2132	&0.5339	&0.3783 & 0.6931 & 0.4303 \\
        \iALS	&0.2106 &0.2106 &\underline{0.5505} &\underline{0.3862} &  \underline{0.7109} &  \underline{0.4382} \\   
        NMF 	&0.2056 &0.2056 &0.5171 &0.3651 & 0.6844 & 0.4192 \\   
        NeuMF	&0.2088 &0.2088 &0.5411 &0.3803 & 0.7093 & 0.4349 \\   
    	\midrule 
    	AE & 0.0975 & 0.0975 & 0.3101 & 0.2064 & 0.4685 & 0.2578\\
        HAE(H, $1$) & 0.2091 & 0.2091 & 0.5247 & 0.3729 & 0.6780 & 0.4231\\
        HAE(M, $1$) & 0.2086 & 0.2086 & 0.5278 & 0.3747 & 0.6818 & 0.4251 \\
        HAE(H, $0.0016$) & 0.2003 & 0.2003 & 0.5285 & 0.3709 & 0.6995 & 0.4269 \\
        HAE(M, $0.0016$) & 0.2108 & 0.2108 & 0.5480 & 0.3845 & 0.7051 & 0.4360 \\
        \midrule
        \textbf{H-VAE}($0.0016$) & \underline{0.2177} & \underline{0.2177} & 0.5400 & 0.3835 & 0.7023 & 0.4366	\\
        \bottomrule
    \end{tabular}
\end{table}

\endgroup

\section{Results}
Our first goal is to analyze how hyperbolicity affects the learning ability of models with respect to the provided objective function. We use validation data for visualizing the difference. As it can be clearly seen from Figure \ref{fig:hyp_comp}, there is a dramatic difference between Euclidean and non-Euclidean geometry models. The Euclidean one fails to find a good configuration during the validation phase, which is indicated by a much lower position of the ``box-plot''. In contrast, both \texttt{HypLinear} and \texttt{M\"obiusLinear}-based networks achieve much higher prediction quality, with the latter typically getting slightly better maximum scores. Morevover, our models tend to be more stable against variations in hyperparameters configuration and only require a few epochs for achieving close to the maximum score (the latter is omitted from graphs due to space constraints). Figure \ref{fig:batch} also shows that pre-estimating space curvature parameter $c$ stabily improves prediction quality of the models, even though the effect is not always very pronounced.

In the second group of experiments, we provide the test results for all our models, including the variational ones. We compare them with common baselines and recent state-of-the-art analyzed in \cite{dacrema2019worrying,dacrema2019troubling}. Tables \ref{tab:Mult-VAE-results} and \ref{tab:NCF-results} consolidate our findings. Remarkably, across all datasets, our shallow variational hyperbolic autoencoder stays close to or even outperforms other competitors at least in one target metric. The results for other metrics at different values of top-$n$ are slightly worse. Nevertheless, considering that we optimize a trivial objective function and no elaborate fine-tuning scheme is employed, our approach opens many opportunities for further improvements. 

\section{Discussion and conclusion}
We presented a hyperbolic autoencoder approach for the standard task of collaborative filtering. The proposed solution utilizes only a single hidden layer yet is capable of competing with modern state-of-the-art models. Our approach has a clear advantage of not introducing excessive complexity, which not only makes it lightweight but also simplifies its analysis. The hyperbolic models based on our approach efficiently capture the underlying structure of data, which immediately leads to a boost in prediction quality. The models exhibit better optimization stability by being less sensitive to a bad choice of hyperparameters. The minimalistic structure of our solution remains flexible and can be further improved with domain-specific components and better optimization objectives. However, care must be taken in order to ensure that the structural properties of the non-euclidean geometry are not violated in the extended solutions, which presents an interesting direction for further research.


\begin{acks}
Authors are thankful to the Sberbank Artificial Intelligence Laboratory for the provided funding of the work.
\end{acks}

\bibliographystyle{ACM-Reference-Format}
\bibliography{biblio}

\end{document}